# Tuning the tilting of the spiral plane by Mn doping in YBaCuFeO$_5$ multiferroic

X. Zhang[1], A. Romaguera[1], O. Fabelo[2], F. Fauth[3], J. Herrero-Martín[3] and J.L. García-Muñoz[1*]

[1]*Institut de Ciència de Materials de Barcelona, ICMAB-CSIC, Campus UAB, 08193 Bellaterra, Spain*
[2]*ILL-Institut Laue Langevin, 38042 Grenoble Cedex, France.*
[3]*CELLS-ALBA Synchrotron, 08290 Cerdanyola del Vallès, Barcelona, Spain.*

**\* Corresponding autor:**
Prof, José Luis García-Muñoz
*E-mail:garcia.munoz@icmab.es*
Phone number: +34 - 935801853

*Institut de Ciència de Materials de Barcelona, ICMAB-CSIC, Campus UAB, 08193 Bellaterra, Spain*

**Abstract**

The layered perovskite YBaCuFeO$_5$ (YBCFO) is considered one of the best candidates to high-temperature chiral multiferroics with strong magnetoelectric coupling. In RBaCuFeO$_5$ perovskites (R: rare-earth or Y) A-site cations are fully ordered whereas their magnetic properties strongly depend on the preparation process. They exhibit partial cationic disorder at the B-site that generates a magnetic spiral stabilized through directionally assisted long range coupling between canted locally frustrated spins. Moreover the orientation of its magnetic spiral can be critical for the magnetoelectric response of this chiral magnetic oxide. We have synthesized and studied YBaCuFe$_{1-x}$Mn$_x$O$_5$ samples doped with Mn, with the aim of increasing spin-orbit coupling effects, and found that the overall Fe/Cu cation disorder at the B-sites can be increased by doping without changing the sample preparation process. In YBaCuFe$_{1-x}$Mn$_x$O$_5$ samples prepared under the same conditions, the T-x magnetic phase diagram have been constructed in the range 10K-500K combining magnetometry, X-ray and neutron powder diffraction measurements. The tilting angles of the spins in the collinear, $\theta_{col}$, and spiral phases, $\theta_{spiral}$, barely vary with temperature. In the collinear phase $\theta_{col}$ is also independent of the Mn content. In contrast, the presence of Mn produces a progressive reorientation of the plane of the magnetic helix in the incommensurate phase, capable to transform the helicoidal spin ordering into a cycloidal one, which may critically determine the ferroelectric and magnetoelectric behavior in these compounds. Some of the observations are of interest for engineering and developing this family of potential high-temperature multiferroics.

## 1. Introduction

Frustration, or the inability to satisfy all interactions, is at the origin of attractive phenomena and properties in complex magnetic materials. The discovery of new classes of frustrated materials where the charge, magnetic or elastic orders and the (ferro-)electric properties are strongly coupled (improper multiferroics, MF) is generating a flurry of activity [1]. However, the low-magnetic ordering temperatures usually found in frustrated magnets (typically <100 K) critically restrict the potential uses of magnetoelectric multiferroics for spintronics and low-power magnetoelectric devices.

The frustrated multiferroic candidate $YBaCuFeO_5$ (YBCFO) is considered one of the most motivating exceptions [2–11]. It represents a fascinating reference of a magnetic lattice without geometric frustration where competing nearest-neighbor interactions can stabilize a ferroelectric spiral phase up to high temperatures. A commensurate (CM) $\mathbf{k}_1=(1/2,1/2,1/2)$ antiferromagnetic (AF) phase below $T_{N1} \sim$ 440K is followed by an incommensurate (ICM) spiral order ($\mathbf{k}_2 = (1/2,1/2,1/2 \pm q)$, $T_{N2} \sim$ 230 K) [2–4]. Not only YBCFO has one of the highest critical temperatures among the magnetically driven multiferroics, but most encouraging, this perovskite exhibits an extraordinary tunability of its spiral ordering temperature ($T_{N2}=T_S$). Several recent works have shown that $T_{N2}$ can be increased by more than 200 K, thus reaching values far beyond room temperature: (i) by manipulating the Cu/Fe chemical disorder in the bipyramids (Morin et al, [5]) and (ii) by chemical pressure (T. Shang et al [6]). The chemical disorder in the YBCFO structure is associated to $Fe^{3+}/Cu^{2+}$ occupational disorder, which can change with the preparation method and the nature of the samples (powders, films, single crystals, etc.). It is a crucial ingredient since a certain degree of chemical disorder was found essential for stabilizing the ICM magnetic phase [4,5,7,8].

The importance of the Cu/Fe chemical disorder in the structure is associated to its impact on the magnetic interactions and the generation of frustration. In this regard, the sign and strength of the main nearest-neighbours (NN) exchange interactions depend on the spatial distribution of $Fe^{3+}$ and $Cu^{2+}$ ions [4]. Parallel to the *c* axis (in addition to the inter-bipyramids AFM $J_{c1}$ exchange), $J_{c2}$ within a bipyramid can be of different sign and magnitude: (i) the strong AFM interaction between two Fe ions, (ii) the FM exchange between a Fe-Cu pair or (iii) the weak Cu-Cu exchange. Perpendicular to the *c* axis, $J_{ab}$ AFM exchanges are due to NN Cu-Cu, Fe-Cu and Fe-Fe pairs in the *ab* plane. Therefore, the FM exchange between a Fe-Cu pair in a bipyramid turns into a strong AFM Fe-O-Fe coupling if the bipyramid is occupied by two $Fe^{3+}$ ions, generating frustration.

In Y(Ba,Sr)$CuFeO_5$ and R(Ba,Sr)$CuFeO_5$ oxides (R: rare-earth or Y), where the cationic substitution affects the A site of the perovskite, the interatomic distances between the magnetic transition metal atoms (TM) are modified respect to YBCFO [6]. Of great importance are the changes in the distances parallel to c as they can tune the magnitude of the competing interactions in that direction [7,8]. A shrinkage in the height of the bipyramids ($d_2$ distance, e.g. by partially substituting $Ba^{2+}$ by the smaller $Sr^{2+}$) or an increase of the distance $d_1$ between bipyramids (e.g. by substituting $Y^{3+}$ by bigger trivalent rare-earth) produce the rise of the spiral transition temperature [6]. The principal theoretical model justifying the onset of the spiral phase in this structure (and probably in other compounds with magnetic spiral order not well understood before) has been developed by Scaramucci *et al*. in refs. [7] and [8]. In a non-frustrated lattice of Heisenberg spins (perfect Cu/Fe order) with only NN interactions, chemical disorder and strong AFM Fe/Fe bonds in bipyramids are introduced as dilute impurity bonds that substitute weak FM Fe/Cu bonds and generate frustration within the bilayers. For sufficiently low dilution, the induced frustration results in a local canting of the spins associated to the impurity bond. Thanks to the identical orientation of the impurity bonds (always parallel to the *c* axis ) a magnetic spiral can be stabilized through long range coupling between the local cantings. A continuous twist can be induced parallel to their direction (when all the impurity bonds point along to a single crystallographic direction), favoured in Bravais lattices in which neighbouring impurities are closer in the plane than along the orientation of the impurity bonds [7,8]. Essential ingredients for this mechanism are the sign and magnitude of the main nearest-neighbours (NN) exchange interactions (along the *c* axis and in the *ab* plane).

In the model spin-orbit interactions play a key active role for determining the common plane of the NN spins along-*c* that become coplanar [7]. Hence, spin-orbit coupling effects would be very relevant, the most important likely being in the form of Dzyaloshinskii–Moriya (DM) terms [12,13]. Considering the form of the DM terms and the magnetic wave vector parallel to **c**, a helical type arrangement of the spins parallel to the **ab** plane would not bring about spontaneous polarization in YBCFO. This has been the cause alluded to explain the absence of ferroelectricity in the YBCFO single crystal studied in ref. [9]. In the work by Lai *et al*. the magnetic model refined from neutron diffraction patterns, collected using parts of the crystal crushed into powder form, shows a spiral with the rotation plane parallel to the **ab** plane (helix). A tendency to spirals with helical orientation in YBCFO was also proposed in ref. [14] using DFT calculations within LSDA+U+SO approximation.

Therefore, of great importance is that despite the fact that some compositions present the spiral transition well beyond room temperature (with Ts values up to near 400 K [6]), spontaneous ferroelectricity above room-temperature has not been observed or confirmed in these samples with high Ts values [14,15]. Given that the number of experimental reports on single-crystals are yet extremely scarce, further neutron and polarization characterization of quality crystals is needed to explain the frequent absence of macroscopic polarization in YBCFO samples.

There is presently a need for identifying new strategies that can optimise and upgrade the magnetic and magnetoelectric properties of these spiral ferroelectrics. In this structure based on bipyramidal layers the partial substitution of $Fe^{3+}$ by less symmetric $Mn^{3+}$ ions in $FeO_5$ pyramids offers new possibilities for engineering its high-temperature response that need to be explored. This work presents a selective substitutional study of the $FeO_5$ pyramids in YBCFO high-temperature multiferroic candidate. Thanks to the similarity of $Mn^{3+}$ and $Fe^{3+}$ ionic sizes (both ≈0.58 Å in pyramidal coordination [16]), in $YBaCu(Fe_{1-x}Mn_x)O_5$ one could increase the spin-orbit coupling through the partial substitution of highly symmetric $Fe^{3+}$ ions ($3d^5$, L=0, S=5/2) by $3d^4$ $Mn^{3+}$ ions (L=2, S=2). In samples prepared with identical cooling rates, the concentration of defects associated to Fe/Cu cation disorder are found to increase and can be tailored by adding a third metal to the B-site. Most interestingly, in this structure the tilt of the rotation plane of the spins in the spiral phase can be tuned. When increasing the Mn content it moves away from the **ab** plane where the DM-based models predict null spontaneous polarization.

**2. Experimental Section**

Doping YBCFO with Mn ions was attempted at the two metal sites: $YBaCu(Fe_{1-x}Mn_x)O_5$ ($Fe^{3+}\leftrightarrow Mn^{3+}$) and $YBa(Cu_{1-x}Mn_x)FeO_5$ ($Cu^{2+}\leftrightarrow Mn^{2+}$) by solid-state reaction and sol-gel methods. Only samples of the first type were successfully prepared. The tetragonal *P4mm* structure formed by bipyramidal layers could not be stabilized when Mn tries to get introduced substituting the divalent Cu sites to form $YBa(Cu_{1-x}Mn_x)FeO_5$, clearly indicating a strong preference of Mn for the trivalent Fe sites in the structure. So, $YBaCu(Fe_{1-x}Mn_x)O_5$ samples with compositions x=0, 0.01, 0.05, 0.10, 0.15 and 0.20 were successfully prepared by solid-state reaction. Stoichiometric amounts of the high purity precursors $Y_2O_3$, $BaCO_3$, CuO, $Fe_2O_3$ and $Mn_2O_3$ were used. After a pre-annealing of $Y_2O_3$ oxide at 900 °C for 10h, all precursors were thoroughly mixed, grounded in an automatic agate mortar, pressed into pellets and then sintered at 1100 °C for 50 h in air. Finally the samples were cooled down to room temperature inside an oven at a controlled rate of 300 K/h. For the purpose of comparing their physical properties it is important to highlight that the same cooling procedure was applied to all compositions. The samples were first characterized by laboratory X-ray powder diffraction using a Siemens D-5000 diffractometer ($\lambda$ [Cu K$_\alpha$]= 1.54 Å). Macroscopic magnetic measurements were performed using a superconducting Quantum Interference Magnetometer Device (SQUID, Quantum Design Inc) for recording magnetic data below room-temperature (RT). In addition, the temperature dependence of the magnetization (*M-T*) was recorded in the interval 300-650 K using a vibrating sample magnetometer (VSM) in a Physical Properties Measurement System (PPMS, Quantum Design Inc). Synchrotron X-ray powder diffraction measurements were performed at 300 K on the BL04-MSPD beamline[17] of the ALBA Synchrotron Light Facility (Barcelona, Spain) using the position sensitive detector MYTHEN. The wavelength, $\lambda$ = 0.41338(3) Å, was determined by measuring a NIST standard silicon. The samples were loaded in borosilicate glass capillaries (diameter of 0.7 mm) and kept spinning during data acquisition. Neutron diffraction experiments were carried out at the high-flux reactor of the Institut Laue Langevin (Grenoble, France) using the high intensity diffractometer D1B ($\lambda$ = 2.52 Å). Neutron powder diffraction (NPD) patterns were collected at selected temperatures and in continuous mode following temperature ramps within the range 10 to 500K. The applied heating rate was 3K/min. One pattern of YBCFO at 300 K was additionally measured on D2B ($\lambda$ = 1.594 Å). Structural and magnetic Rietveld refinements were carried out using the Fullprof program[18].

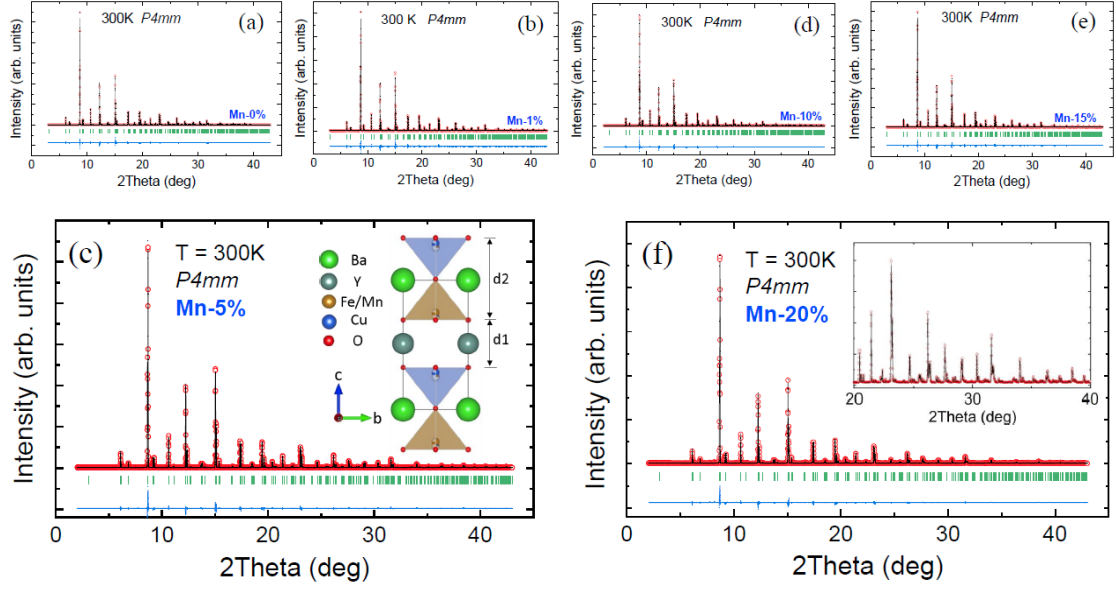

**Figure 1.** Rietveld refinement (black curve) of the synchrotron X-ray intensities (red circles) at 300 K. Bottom blue line is the observed-calculated difference. (a) Mn-0%. (b) Mn-1%. (c) Mn-5%. Inset: projection of the *P4mm* structure showing the d1 (inter-bilayers) and d2 (bypyramid layer) distances. (d) Mn-10%. (e) Mn-15%. (f) Mn-20%. Inset: Detail of the high-angles region.

## 3. Results and discussion

### 3.1 Structural characterization

Figure 1 displays the refined synchrotron X-ray diffraction patterns (SXRD) of the six YBaCuFe$_{1-x}$Mn$_x$O$_5$ samples investigated. Patterns were similar for all compositions, hereafter denoted as Mn-0%, Mn-1%, Mn-5%, Mn-10%, Mn-15% and Mn-20%. From SXRD patterns, secondary phases such as Y$_2$O$_3$ or CuO were detected up to ~0.2% wgt (wgt: weight) in most cases and below ~0.9% wgt in the most heavily doped samples. Samples are well described using the *P4mm* symmetry, which gives the best agreement factors and accounts for different spatial disorder of the TMs [4–6,19].

**Table I**. Structural parameters at T=300K and agreement factors from Rietveld refinement of MSPD@Alba synchrotron data ($\lambda$=0.41338(3) Å). (*: minority fraction; +: occupation of the majority cation in each pyramid). The coordinates of the two positions (M1 and M2) for each metal are related by z(M1)+ z(M2)=1.

| YBaCuFe$_{1-x}$Mn$_x$O$_5$ | x=0 | x=0.01 | x=0.05 | x=0.10 | x=0.15 | x=0.20 |
|---|---|---|---|---|---|---|
| a | 3.87463(3) | 3.87439(3) | 3.87354(3) | 3.87274(4) | 3.87189(4) | 3.87052(5) |
| c | 7.66267(5) | 7.66349(6) | 7.66479(6) | 7.66816(8) | 7.67013(8) | 7.67265(9) |
| v | 115.038(1) | 115.036(1) | 115.005(2) | 115.008(2) | 114.987(2) | 114.943(2) |
| z (Y) (0 0 z) | 0.5097(5) | 0.5096(5) | 0.5116(4) | 0.5105(5) | 0.5036(5) | 0.5031(6) |
| z (Fe1) (½ ½ z) | 0.7540(6) | 0.7507(7) | 0.7551(6) | 0.7538(6) | 0.7543(7) | 0.7524(6) |
| z (Cu1) (½ ½ z)* | 0.7145(3) | 0.7156(3) | 0.7150(3) | 0.7153(2) | 0.7184(4) | 0.7162(3) |
| z (Cu2) (½ ½ z) | 0.2855(3) | 0.2844(3) | 0.2850(3) | 0.2847(2) | 0.2816(4) | 0.2838(3) |
| z (Fe2) (½ ½ z)* | 0.2460(6) | 0.2493(7) | 0.2449(6) | 0.2462(6) | 0.2457(7) | 0.2476(6) |
| z (O$_1$) (½ ½ z) | 0.010(3) | 0.008(3) | 0.011(3) | 0.011(3) | 0.009(3) | 0.007(3) |
| z (O$_2$) (0 ½ z) | 0.327(1) | 0.325(1) | 0.327(1) | 0.325(1) | 0.324(1) | 0.320(2) |
| z (O$_3$) (0 ½ z) | 0.693(1) | 0.691(1) | 0.697(1) | 0.696(1) | 0.697(1) | 0.695(2) |
| Occ (majority)+ | 0.772(22) | 0.764(24) | 0.704(22) | 0.688(26) | 0.671(28) | 0.668(28) |
| $\chi^2$ | 50.0 | 67.5 | 70.9 | 57.6 | 61.2 | 60.8 |
| $R_B$ | 6.02 | 5.27 | 5.44 | 4.57 | 3.80 | 4.90 |
| $R_f$ | 7.28 | 6.24 | 7.16 | 5.97 | 4.30 | 5.90 |

Table I reports structural parameters and agreement factors obtained from Rietveld refinement at 300 K as a function of the manganese content. We refined the two positions and the respective fractions of Cu and (Fe,Mn) ions in each pyramid, the upper (brown) and the lower (blue) pyramids within the unit cell (see inset of Fig. 1(a)). For a proper refinement of the positions and the fractions of the metals in the pyramids, the z-coordinates of the two positions that a given metal M (either Fe/Mn or Cu) occupy in the cell (M1 and M2) were refined in the form z(M1)+z(M2)=1. Fig. 2(a) displays the evolution of the refined chemical disorder. Given the importance of how Fe/Cu atoms are distributed throughout the crystal structure, Table I unveils the fraction of Cu atoms that occupy Fe sites and vice versa. A first appealing result in this series prepared using the same cooling rate is that the presence of a third metal (Mn) systematically increases the degree of Fe/Cu chemical disorder. A certain chemical disorder is a prerequisite for the magnetic order in this structure to evolve from the commensurate to the incommensurate phase. So, from the evolution of the occupation at B-sites depicted in Fig. 2(a), one can see that chemical disorder increases from $n_{oc}$~23% without Mn up to $n_{oc}$~33% in the Mn-20% sample. It is worth underlying that this is a remarkable relative raise (~40%) achieved in samples prepared using the same synthesis process and an identical cooling rate of 300 K/h.

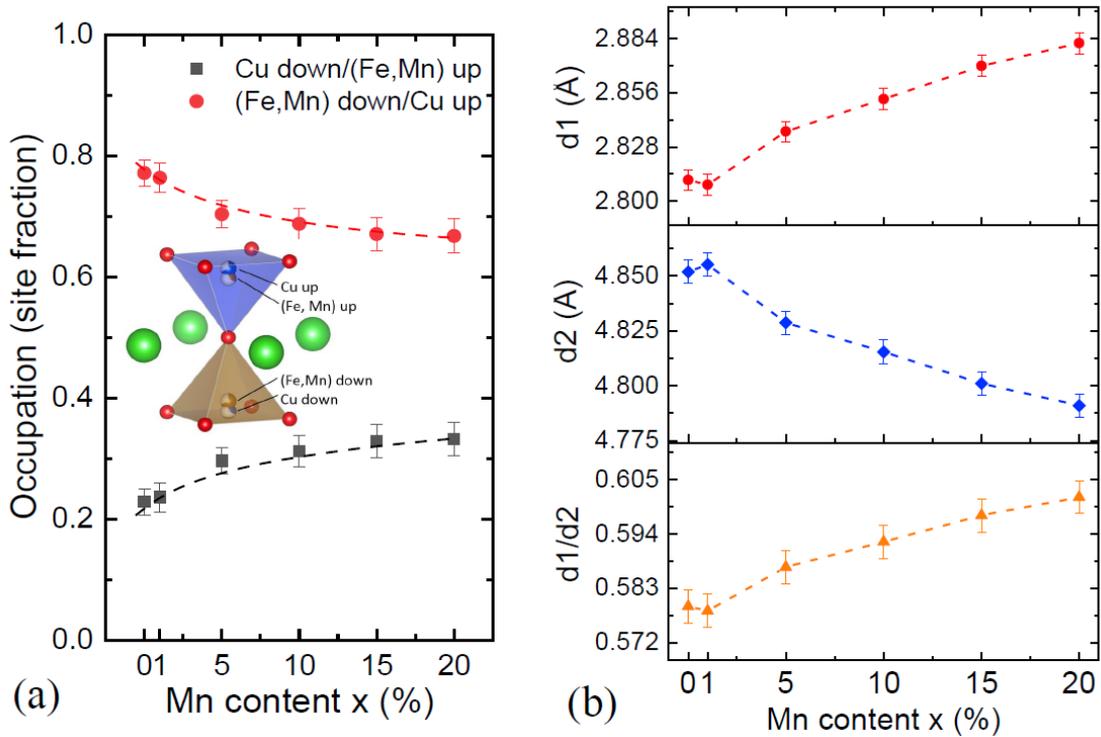

**Figure 2.** (a) Evolution with Mn doping of the average site occupation in the bypiramids of $YBaCuFe_{1-x}Mn_xO_5$ samples prepared following the same cooling process (same rate of 300 K/h). An occupation 0.5 would correspond to a random B-site cation distribution. (b) Evolution of the characteristic distances d1 (separation between bilayers), d2 (thickness of the bilayers) and the ratio d1/d2.

Table S1 (Supplementary Information) compares the structural results of the Rietveld fits for the YBCFO reference sample of this work obtained from high-resolution SXRD and NPD data. Moreover, they are also compared to those reported for the YBCFO sample in ref [4]. Fig. S1 (Supplementary Information) plots the fit of the high-resolution D2B

neutron pattern for our undoped sample. The evolution of the cell parameters and volume included in Table I has been plotted in Fig. S2 (Supplementary Information). Systematic increases of the *c* parameter are observed, in addition to a relaxation of the tensile tetragonal distortion (increase of $c/2a < 1$) and a volume decrease, although the changes are relatively very small. The expansion of the *c* parameter along the z-axis is likewise closely related to the evolution of d1 and d2, the two main distances that characterize the [CuFeO$_9$] bilayers of corner-sharing square pyramids [6]. They are shown in the inset of Fig. 1(a). Distance d1 refers to the separation between bilayers and d2 represents the thickness of each individual bilayer, i.e. the height of the unit formed by two corner-sharing MO$_5$ square pyramids (bipyramid; M: metal). In this structure the height (d2) of the CuFeO$_9$ bipyramids is longer than the distance d1 between them (d1< d2). Fig. 2(b) illustrates the evolution of d1 and d2, showing a monotonous expansion of the separation between bipyramids (d1): $\Delta d1 \approx +0.072$ Å up to 20% Mn. Moreover the thickness of the bipyramid layer (d2) decreases ($\Delta d2 \approx -0.064$ Å up to 20% Mn). The expansion in d1 is more pronounced than the contraction in d2 producing the increase of *c*. The evolution of the ratio d1/d2 is also depicted in Fig. 2(b). Additional information such as the evolution of the main interatomic distances in the structure is given in Fig. 3 and Table S2 of the Supplementary Information.

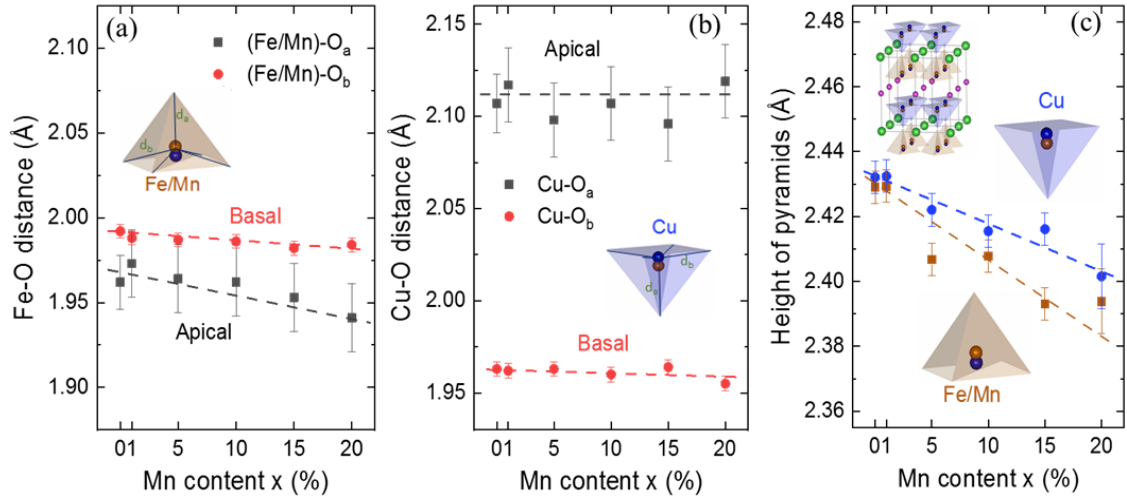

**Figure 3.** Evolution of the (a) (Fe/Mn)-O and (b) Cu-O interatomic distances at room temperature in respectively the upper and lower pyramids of the chemical unit cell. d$_a$='apical' and d$_b$='basal' (equatorial) distances in pyramidal coordination. Insets: upper (brown) and lower (blue) pyramids in the unit cell. (c) Height of the [Fe/Mn]O$_5$ and CuO$_5$ pyramids. Inset: projection of the structure showing the [CuFeO$_9$] bilayers of corner-sharing square pyramids (green atoms: Ba; red: Y). The color of each pyramid corresponds to the color of the dominant cation in it.

A look to the refined interatomic distances reveals very small variations. We can see in Fig. 3(b) that the Jahn-Teller splitting between basal and apical distances around Cu is preserved along the series. No apparent changes are detected in the large difference between the interatomic Cu-Oa (apical) and Cu-Ob (basal) distances of the $CuO_5$ pyramids (the Jahn-Teller distortion around $Cu^{2+}$ sites ($t_{2g}^6 e_g^3$)). It is thus confirmed that the large splitting between shorter Cu-Ob bonds and a much longer apical Cu-Oa distance is not modified. On another hand $Mn^{3+}$ is also a potential Jahn-Teller active ion due to its expected $t_{2g}^3 e_g^1$ electronic configuration. For the case of a $Mn^{3+}O_5$ coordination polyhedron, Millange *et al*. reported the crystal structure of the related oxide $YBaMn_2O_5$ [20]. In this compound, having a similar structure as $YBaCuFeO_5$, the occurrence of $Mn^{2+}/Mn^{3+}$ charge order (perfect order in the A and B sites of the perovskite) results in the presence of two different sites for the two manganese ions and *P4/nmm* symmetry. The bipyramids are formed by one regular and one distorted pyramid. For the pyramidally coordinated $Mn^{3+}$, the $d_{z^2}$ orbital extends along c and the unoccupied $d_{x^2-y^2}$ orbital along [110] and [1-10], producing four basal Mn-Ob distances of 1.917 Å and one apical Mn-Oa distance of 2.048 Å [20]. This large Jahn-Teller splitting would produce an enlargement of the apical Fe-Oa bonds if $Mn^{3+}$ adopted the same electronic configuration in $YBaCuFe_{1-x}Mn_xO_5$. Interestingly, the evolution of the apical and basal distances displayed in Fig. 3(a) strongly suggests that in this series Mn ions do not adopt the Jahn-Teller configuration $t_{2g}^3 e_g^1$ with the $d_{z^2}$ orbital pointing along [001]. The expected enlargement in the apical Fe-Oa distance as a larger fraction of Mn substitutes Fe atoms is at odds with these results. Instead, Fig. 3(a) suggests a slight shrinking in the average Fe-Oa bond of the $(Fe,Mn)O_5$ pyramids when increasing the presence of Mn. Although a variety of orbitally ordered and disordered states have been observed in different manganites containing $Mn^{3+}$ ions during the last decades, a very specific feature in $YBaCuFe_{1-x}Mn_xO_5$ is the concurrence of two potential strong Jahn-Teller-active cations ($Cu^{2+}$ and $Mn^{3+}$) in bipyramids where $CuO_5$ and $MnO_5$ units are sharing the apical oxygen. The evolution of the interatomic distances could indicate that the robust Jahn-Teller distortion confirmed in $CuO_5$ pyramids is impeding the deformation of the $MnO_5$ polyhedra in the $CuMnO_9$ bipyramids. The simultaneous coexistence in a bipyramid of (i) a Jahn-Teller distorted $Cu^{2+}$ pyramid, very elongated, and (ii) a Jahn-Teller $t_{2g}^3 e_g^1$ $Mn^{3+}$ ion with the $d_{z^2}$ orbital pointing parallel to the Cu-Oa bond along z, very likely, has a too high energy penalty. The same conclusion is drawn by the evolution of the heights in the upper (Fe) and lower (Cu) pyramids displayed in Fig. 3(c). Additional spectroscopic studies (preferably on single crystals) are needed to get a more complete description of the local electronic structure in this series.

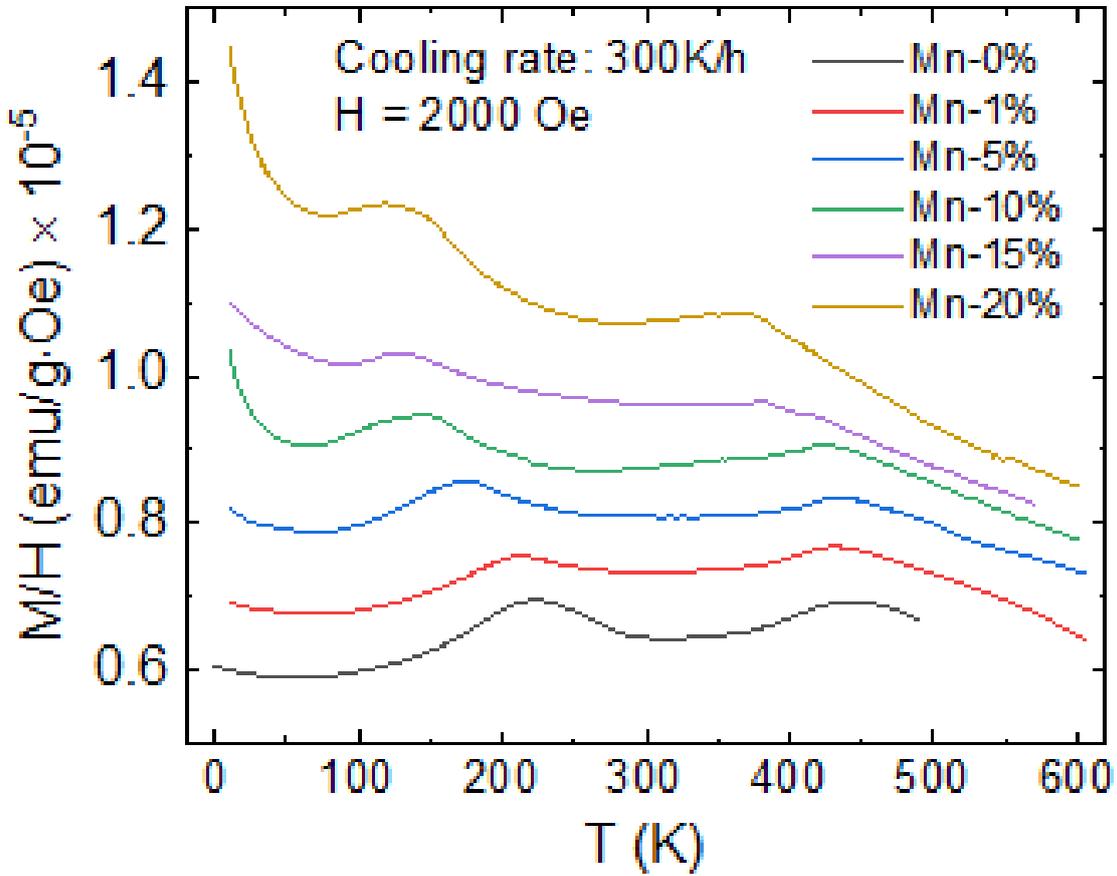

**Figure 4.** Magnetic susceptibility curves (2 kOe) of the $YBaCuFe_{1-x}Mn_xO_5$ samples prepared using identical cooling rates (300 K/h) in the last annealing. The two transitions observed in all the samples correspond to the CM collinear ($T_{N1}$, $k_1$) and ICM ($T_{N2}$, $k_2$) magnetic phases. Successive curves were shifted by $+10^{-6}$ emu g$^{-1}$ Oe$^{-1}$ for clarity.

*3.2 Magnetic transitions*

As expected, the Neel temperatures undergo a certain decrease when reducing the concentration of $Fe^{3+}$ ions in the structure. Susceptibility measurements were performed in a dc magnetic field of 2 kOe after field-cooling (FC) from 5 K to 600K, as shown in Figure 4. Two separated magnetic transitions occur in all compositions, associated respectively to the onset of the CM ($T_{N1}$) and the ICM ($T_{N2}$) magnetic orders. This was confirmed by neutron diffraction.

All compositions were characterized by neutron diffraction in the 10K-500K range, using the D1B diffractometer. Upon cooling from 500K two distinct sets of new magnetic Bragg reflections appear at $T_{N1}$ and $T_{N2}$. First, new peaks indexed as (*h*/2, *k*/2, *l*/2) appear at $T_{N1}$ (>350 K, for all the samples), associated to the propagation vector $k_1$= (1/2, 1/2, 1/2) (A-point of the Brillouin Zone). Upon further cooling new magnetic peaks compatible with the incommensurate propagation vector $\mathbf{k}_2=(1/2, 1/2, 1/2\pm q)$ emerge at $T_{N2}$ (<250 K for all the samples) and persist down to base temperature. The T-x phase diagram for YBaCuFe$_{1-x}$Mn$_x$O$_5$ samples is shown in Fig. 5(a).

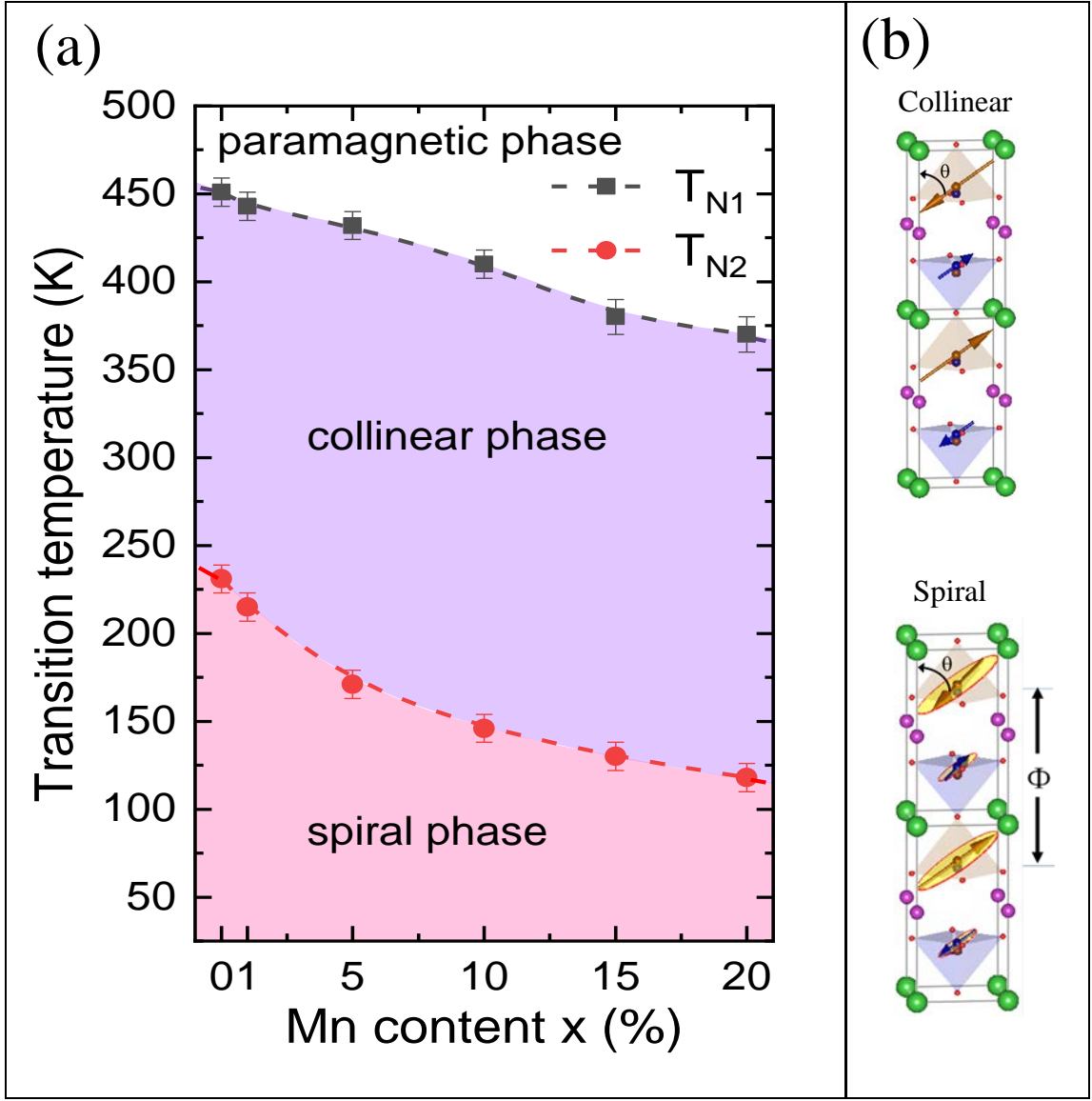

**Figure 5.** (a) T-x phase diagram for YBaCuFe$_{1-x}$Mn$_x$O$_5$ samples. The phase boundaries describe the onset of the magnetic phases according to neutron diffraction data. (b) Sketch of the CM collinear ($T_{N1}$) and ICM spiral ($T_{N2}$) magnetic phases showing the tilting angle $\theta$. The rotation angle $\Phi$ between successive spins along the **c** axis deviates from 180º in the ICM phase. For clarity, the average magnetic moment is depicted in each pyramid using the color of the majority metal.

The partial decrease of $T_{N1}$ is due to a combination of factors: lattice and structural tuning effects, the increase of chemical disorder and also the changes in the NN exchange constants due to the substitution of Fe by Mn atoms. The decrease of the spiral transition temperature ($T_{N2}= T_S$) is not explained by lattice effects or the increase of the Fe/Cu cation disorder that would raise $T_S$. In the light of the model developed by Scaramucci and co-workers [7,8] the decrease should be mainly attributed to the changes in the magnetic exchange couplings parallel to the *c* axis, but not only. Essential ingredients of the model are the exchange interactions within the bipyramids. So, a fraction of the strong AFM $Fe^{3+}$-O-$Fe^{3+}$ bonds (the strongest coupling) would be substituted by the much weaker $Fe^{3+}$-O-$Mn^{3+}$ (or $Mn^{3+}$-O-$Mn^{3+}$) AFM coupling. In the mentioned theoretical Heisenberg model the presence of a fraction of very strong AFM $Fe^{3+}$-O-$Fe^{3+}$ bonds is required. They act as dilute impurity bonds in that generate enough frustration to induce a local canting of the FM spin pairs in the $Fe^{3+}$-$Cu^{3+}$ bipyramids. If instead of a FM Fe-Cu pair, the bipyramid is formed by a $Mn^{3+}$-$Cu^{3+}$ pair (with, respectively, single and double $d_z^2$ orbital occupation) the Goodenough-Kanamori-Anderson (GKA) rules [21–23] also predict ferromagnetic exchange between them. Instead, an AFM coupling is predicted if the single $e_g$ occupation in $Mn^{3+}$ ions would occur in $d_{x^2-y^2}$ orbitals. In the model the long range coupling between locally-canted spins takes place favored by the orientation of these impurity bonds always parallel to the *c* axis [7,8].

On the other hand, some of the pristine $J_{ab}$ antiferromagnetic exchanges (Cu-Cu, Fe-Cu and Fe-Fe pairs) in the **ab** plane would probably be substituted by ferromagnetic couplings, associated to $Cu^{2+}$-$Mn^{3+}$ or $Fe^{3+}$-$Mn^{3+}$ NNs pairs in the plane, thus disturbing the AFM correlations perpendicular to *c*. It should also be pointed out that our structural results are at odds with the expected evolution for the usual Jahn-Teller configuration in pyramidally coordinated $Mn^{3+}$ (with the $d_z^2$ orbital pointing along the vertical axis of the pyramid). Therefore, further investigations of the local electronic structure would be desirable to unambiguously determine some of the actual new exchange couplings associated to the presence of Mn.

Figure 6 compares the low-angle region of the neutron patterns collected at 10 K for all compositions. A systematic evolution is observed when increasing the amount of Mn substitution. As Mn replaces Fe both the ICM and CM magnetic phases coexist, but the relative weight of the latter grows with x. Thus, in Mn-20% both phases still coexist but the CM phase is largely dominant over the ICM one. This evolution in the relative intensities of the magnetic peaks associated to the CM and the ICM propagation vectors indicates that they are due to two separate magnetic phases (not a multi-*k* structure).

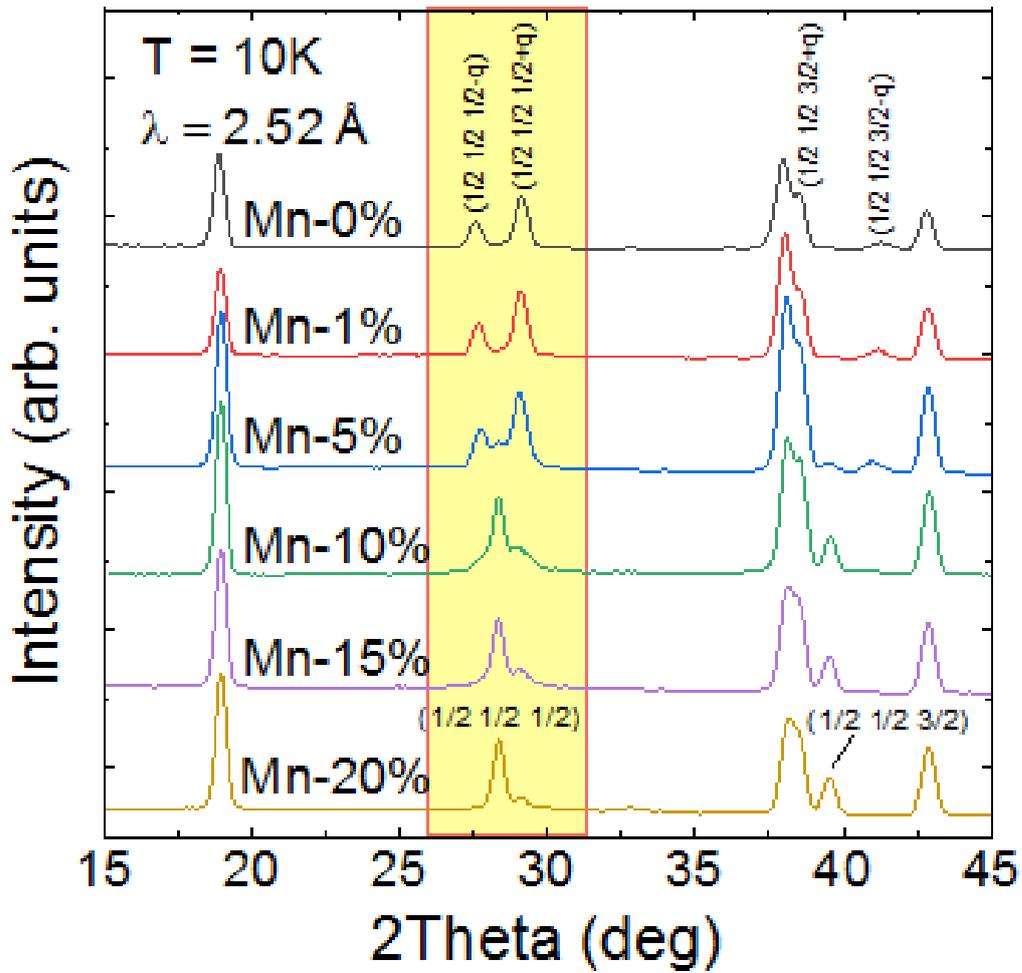

**Figure 6.** Low-angle region of the neutron diffraction patterns recorded at 10 K (d1b@ILL, $\lambda$=2.52 Å) for samples with increasing levels of Mn content.

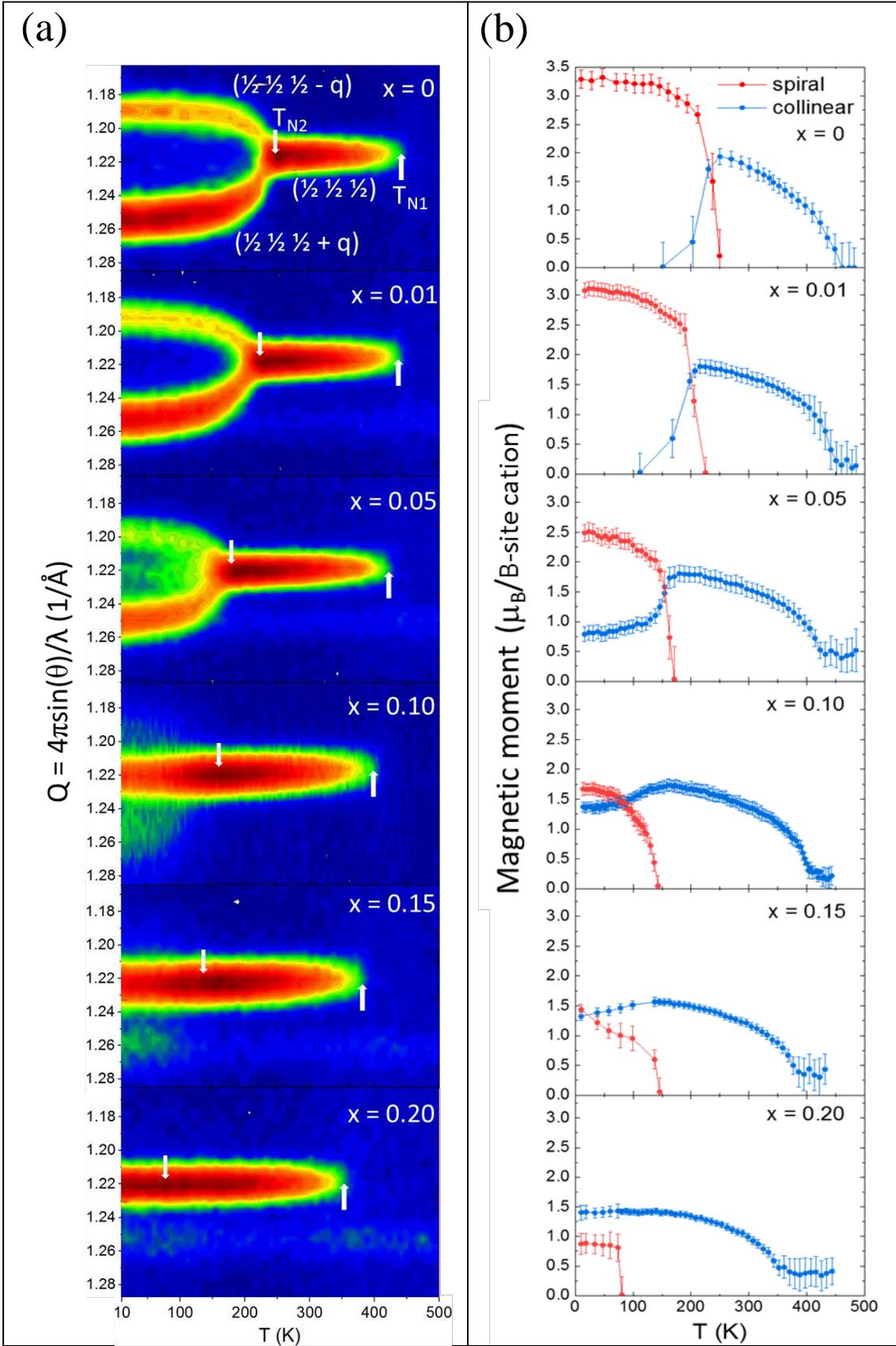

**Figure 7.** (a) T−Q projection of the temperature dependence for the neutron-diffracted intensities around (½ ½ ½) reflection. Evolution with the Mn content (d1b@ILL, $\lambda$=2.52 Å). (b) T-dependence of the (average) ordered magnetic moments associated to the CM and ICM phases.

*3.2.1 Temperature dependence*

NPD patterns were collected using the continuous mode in the high intensity diffractometer D1B, with excellent resolution at low-q and a high efficiency position sensitive detector covering the angular range up to 129º. Around 50 patterns were recorded in the range 10 K-450 K, with a temperature step of 9 K between successive patterns. Figure 7(a) plots a T−Q projection of the neutron-diffracted intensities in a selected angular range, around the (1/2 1/2 1/2) reflection. The evolution of the main magnetic reflections is shown between 10 K and 500 K, illustrating the onset and evolution of the two magnetic phases. The splitting of the $\mathbf{k}_1$=(1/2, 1/2, 1/2) reflections into two satellites signals the onset of the ICM modulation associated to the translational symmetry along the *c* axis. The magnetic modulation parameter $q(T)$ appears at $T_{N2}$ and progresses until it reaches its maximum amplitude $q_0$.

Neutron powder diffraction data are compatible with an ICM spiral phase below $T_{N2}$:

$$\boldsymbol{m}_{lj}(\boldsymbol{k}) = M_R(\boldsymbol{m})(\boldsymbol{u}_j)\cos(2\pi\{\boldsymbol{k}.\boldsymbol{R}_l + \Phi_J\}) + M_I(\boldsymbol{m})(\boldsymbol{v}_j)\sin(2\pi\{\boldsymbol{k}.\boldsymbol{R}_l + \Phi_J\}) \quad \text{(eq. [1])}$$

where $\boldsymbol{m}_{lj}$ is the magnetic moment of the atom *j* in the unit cell *l*, $\boldsymbol{R}_l$ is the vector joining the arbitrary origin to the origin of unit cell *l*, and $\Phi_j$ is a magnetic phase. $\boldsymbol{u}_j$ and $\boldsymbol{v}_j$ in eq. [1] define the orientation of the two perpendicular axes that fix the plane of the helix, where $M_R$ (real) and $M_I$ (imaginary) amplitudes determine the ellipse that envelops the magnetic moments. Due to the intrinsic limitations of the data collected on powder samples (distinct magnetic peaks can share the magnitude of momentum transfer wave vector) it is not possible to independently refine parameters such as the real and imaginary amplitudes ($M_R$ and $M_I$), forcing us to fix their ratio using constraints. The moments at the $Cu^{2+}$ and $Fe^{3+}$ sites of the cell were restricted to the same inclination angle $\theta$ in the neutron refinements. The phase difference φ between the magnetic moments at the two sites was fixed to 180º as found in the YBCFO single-crystal study of ref. [9], and in earlier refs. [3–5]. The relative intensities of the two satellites (1/2 1/2 1/2±q) in Fig. 6 indicate that φ does not deviate from 180º by doping [4,9]. In the Fig. 7(b) we have plotted the refined (average) magnetic moments for each one of these phases and their evolution up to 500K. In these figures the ICM and CM magnetic moments are referred to the full amount of sample. Moreover, their values correspond to ordered moments averaged over Fe and Cu sites, and (for the spiral phase) in the limit $M_R \gg M_I$.

Regarding the magnetic moments the neutron patterns were refined within two imposed limits: (i) m(Fe)=5m(Cu) (as the ratio of their respective unpaired spins), for the lowest-disorder case; and (ii) m(Fe)=m(Cu), for the highest-disorder limit. In addition, for the ICM/spiral phase we considered two limits in the refinements: the limit $M_R \gg M_I$ (high eccentricity) and the circular ellipse model ($M_R = M_I$). Similar agreement factors were obtained in all cases.

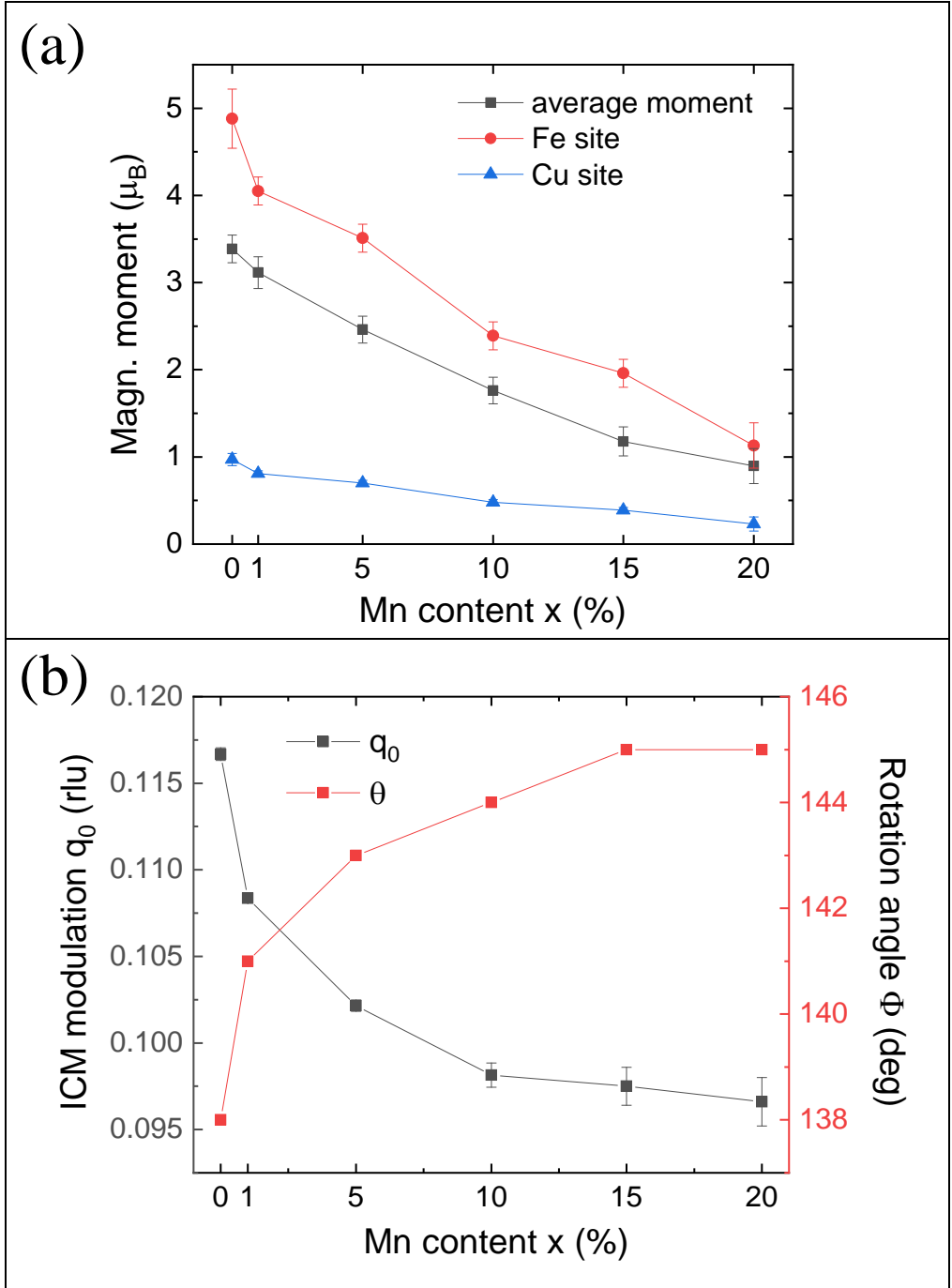

**Figure 8.** Evolution of the ICM magnetic phase with Mn content. (a) Ordered magnetic moments at 10 K (referred to the full sample and $M_R \gg M_I$). Blue and red dots correspond to the case where a constraint in the form m(Fe)=5m(Cu) gets applied. Black dots correspond to the case where the same average moment is refined at Cu and Fe sites; (b) ICM $q_0$ modulation component and associated spin rotation angle Phi (Φ) between successive bilayers.

The evolution of the ordered moments with the fraction of Mn shown in Table II and Fig.8(a) reveals that the ordered moment at Fe/Mn sites decreases faster than expected when substituting S=5/2 $Fe^{3+}$ spins by S=2 $Mn^{3+}$ spins. Moreover, it neither can be ascribed to the increase of chemical disorder between divalent and trivalent ions. Instead, the observed evolution should be attributed to enhanced magnetic frustration by the presence of Mn ions. The magnitude and the sign of some of the pristine interactions should change when $Fe^{3+}$ is substituted by $Mn^{3+}$ increasing frustration. So, for example, the GKA rules for the $Mn^{3+}$/$Cu^{2+}$ and $Mn^{3+}$/$Fe^{3+}$ pairs of NNs in this square-pyramidal arrangement predict FM couplings within the *ab* plane. This is in contrast to the pristine compound (YBCFO), where NN couplings in that plane are always AFM, independently of the $Fe^{3+}$/$Cu^{2+}$ distribution. Within a bipyramid formed by two elongated $Cu^{2+}$ ($3d^4$) and $Mn^{3+}$ ($3d^9$) pyramids (with the occupied $d_{z2}$ orbital along [001]) one would expect FM coupling along **c**. However, the formation of such Jahn-Teller Cu-Mn bipyramids is presumably energetically more expensive than other configurations.

In addition, the evolution with the Mn content of the incommensurability parameter $q_0$ is plotted in Fig. 8(b). $q_0$ corresponds to the maximum amplitude of the ICM modulation. The amplitude $q_0$ of the incommensurability is a measurement of the magnetic rotation of the spin moments between two successive bilayers or, in other words, between two successive unit cells along the *c* axis [see Fig. 5(b)]. In the same plot of the Fig. 8(b) we show the Φ values as determined for the different compositions. The $q_0$ ($x_{Mn}$) evolution is clearly nonlinear and reaches a kind of asymptotic or constant value for x>10% Mn of $q_0$≈0.098 rlu (Φ=215.3º). This value is far from the commensurate limit. Therefore this evolution suggest that increasing x further than 10% Mn produces a reduction of the ICM phase in favor of the CM collinear order, whose spatial distribution is related or conditioned by the local distribution of Mn ions.

*3.2.2 Easy axis and magnetic plane inclination*

The angle $\theta$ represents the angular distance (tilting) between the *c* axis and the direction of the spins, namely (i) the collinear spins direction **u** in the CM phase (T< $T_{N1}$) or (ii) the rotation plane of the helix in the spiral ICM magnetic phase (**u**-**v** plane at T<$T_{N2}$). Due to the tetragonal symmetry the orientation of the moments in the *ab* plane cannot be determined from neutron diffraction data on powder samples. Hence, for simplicity, the director vector **u** can be taken as within the *ac*-plane, **v** being thus parallel to the *b* axis. A $\theta$ value close to 0º means that the easy-axis is close to the *c* axis, whereas for $\theta$ ≈90º the easy-axis is within the **ab-**plane. In the *P4mm* structure when $\theta$ adopts intermediate values (different to 0 or 90º) the magnetic arrangement according to **k**$_1$ or **k**$_2$ translational symmetries requires the concurrent activation of magnetic modes belonging to distinct irreducible representations (magnetic *irreps*) [4]. The temperature evolution of the tilting angle $\theta$ was refined for all the samples between 10K and the onset of the paramagnetic phase using the ratio m(Fe)=5m(Cu). Below T$_{N2}$ we used the $M_R \gg M_I$ limit that allows us to minimize the imprecision (errors) in the determination of $\theta$, maximizing accuracy. The full diffraction patterns were analyzed for all temperature values, using different magnetic phases for the sequential refinements above and below T$_{N2}$, respectively. Initially we used the structures showed in Table I and the six free structural coordinates in the structure ('z' coordinates) were refined as a function of temperature (in addition to temperature factors, cell dimensions, magnetic phases, etc..) but they did not show any thermal evolution and remained constant. The extracted magnetic parameters showed a good convergence and their refined values were the same both refining and keeping invariant the structural z-coordinates. Figure 9 illustrates the results obtained for the six different compositions of YBaCuFe$_{1-x}$Mn$_x$O$_5$ studied.

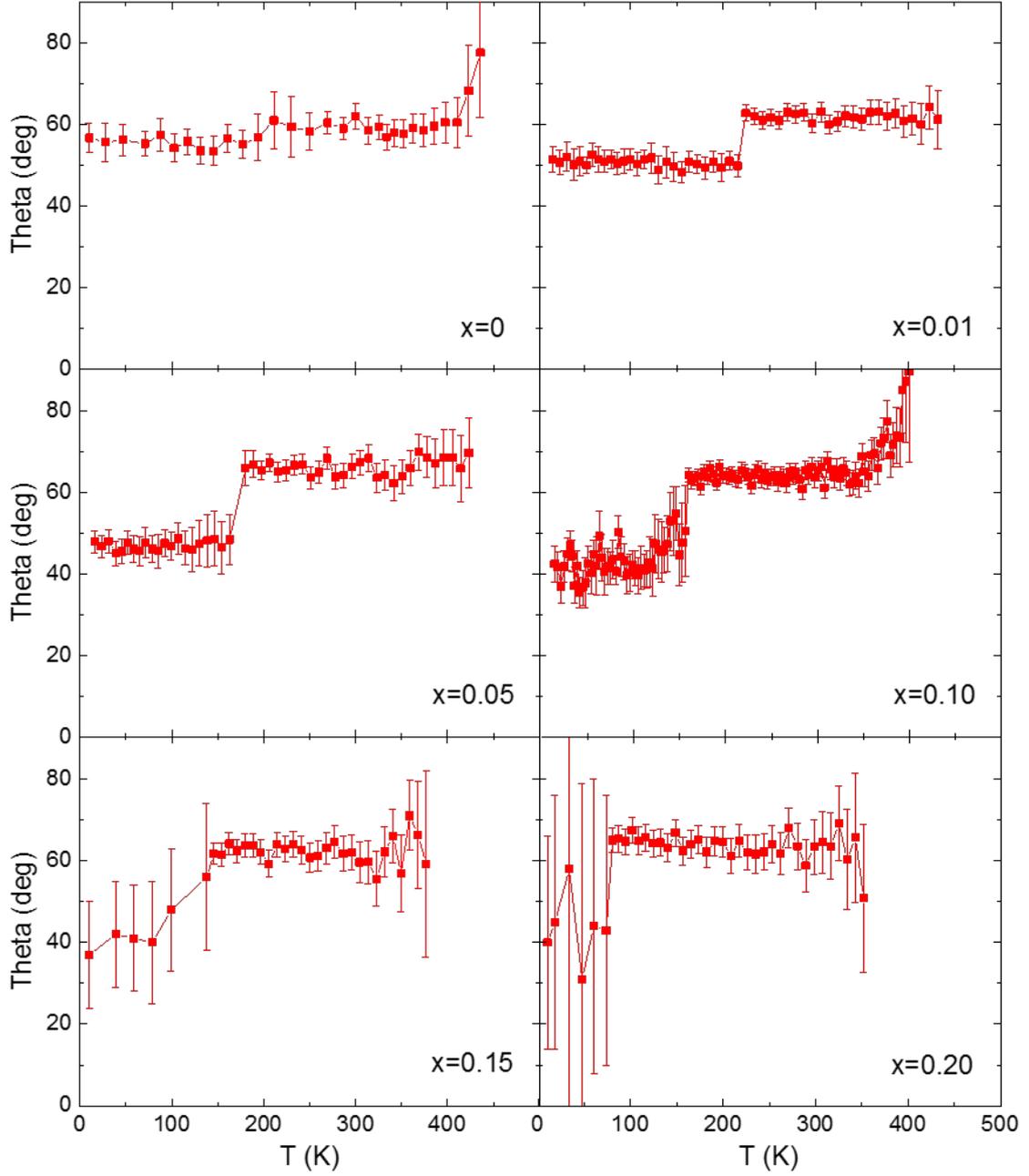

**Figure 9.** Temperature evolution of the inclination angle $\theta$ of the collinear spins ( $T_{N2} < T < T_{N1}$) and of the helix rotation plane (T< $T_{N2}$) in YBaCuFe$_{1-x}$Mn$_x$O$_5$ samples for the different Mn contents.

The evolution of $\theta$(T, x) plotted in Fig. 9 allows us to extract several appealing conclusions. First, within each magnetic regime the tilting $\theta$ remains practically constant. We observe this behavior in all compositions. This result is at odds with some early interpretations of a continuous decrease of $\theta$ when lowering temperature in YBCFO. Second, from present results the orientation of the easy axis is stable in the collinear regime, and equally the tilt of the plane of the helix does not appreciably vary below $T_{N2}$.

However, we observe a sudden fall of the inclination around $T_{N2}$, coinciding with the CM-to-ICM transformation. Indeed Fig. 9 shows a sudden drop of the tilting at the transition in all samples. Third, interestingly, the descent looks more pronounced by increasing the amount of substituted metal. Only in the sample with the highest doping (Mn-20%) the weakness of the ICM reflections precludes unambiguously determining its $\theta$ value.

The refined magnetic moments [m(Fe)=5m(Cu)] and tilting angles $\theta(x)$ obtained refining the CM collinear phase at 300K and the ICM phase at 10K are given in Table II. For the latter, the moments found using the two extreme models Mr>>Mi (*Model1*) and Mr=Mi (*Model2*) are both shown. As expected the refine moments using a sinusoidal modulated order (*Model1*) are greater than those obtained for a circular spiral (*Model2*). The value found at the Fe positions of YBCFO with Model1 at 10 K (4.9(2) $\mu_B$/Fe) is very close to the atomic moment for $Fe^{3+}$ ions. Although this value is too high and hardly compatible with the amount of chemical disorder in the sample, it cannot be taken to rule out Model1 because it results from imposing a constraint over the ordered moments [m(Fe)=5m(Cu)] in the bipyramid that represents the limit of full cationic order. A projection of the refined ICM magnetic order at 10K using Model2 (Mr=Mi, circular spiral) is shown in Figure 10(a) for all doped compositions.

**Table II**. Refined Magnetic moments [imposing m(Fe)=5m(Cu)] and inclination angles ($\theta$) obtained refining the CM collinear phase at 300K and the ICM phase at 10K. At low temperature, the refined moments correspond to models Mr>>Mi (*Model1*) and Mr=Mi (*Model2*).

| sample | 300K (CM) | | 10K (ICM) | | | |
| --- | --- | --- | --- | --- | --- | --- |
| | *Collinear* | | *Model1* (Mi~0) | | *Model2* (Mr=Mi) | |
| | m_Fe ($\mu_B$) | Theta (deg) | m_Fe ($\mu_B$) | Theta (deg) | m_Fe ($\mu_B$) | Theta* (deg) |
| Mn-0% | 2.34 (2) | 59 (2) | 4.9 (2) | 55 (5) | 3.61 (4) | 55 |
| Mn-1% | 2.31 (2) | 62 (2) | 4.1 (2) | 52 (5) | 2.94 (3) | 52 |
| Mn-5% | 2.15 (2) | 66 (2) | 3.5 (2) | 47 (6) | 2.19 (3) | 47 |
| Mn-10% | 1.91 (2) | 64 (2) | 2.4 (2) | 43 (8) | 1.78 (3) | 43 |
| Mn-15% | 1.42 (2) | 65 (3) | 1.9 (2) | 35 (12) | 1.48 (3) | 35 |
| Mn-20% | 1.36 (2) | 65 (4) | 1.1 (2) | 55 (52) | 0.86 (4) | 55 |

\* Theta ($\theta$) in Model2 is fixed to its refined value in Model1.

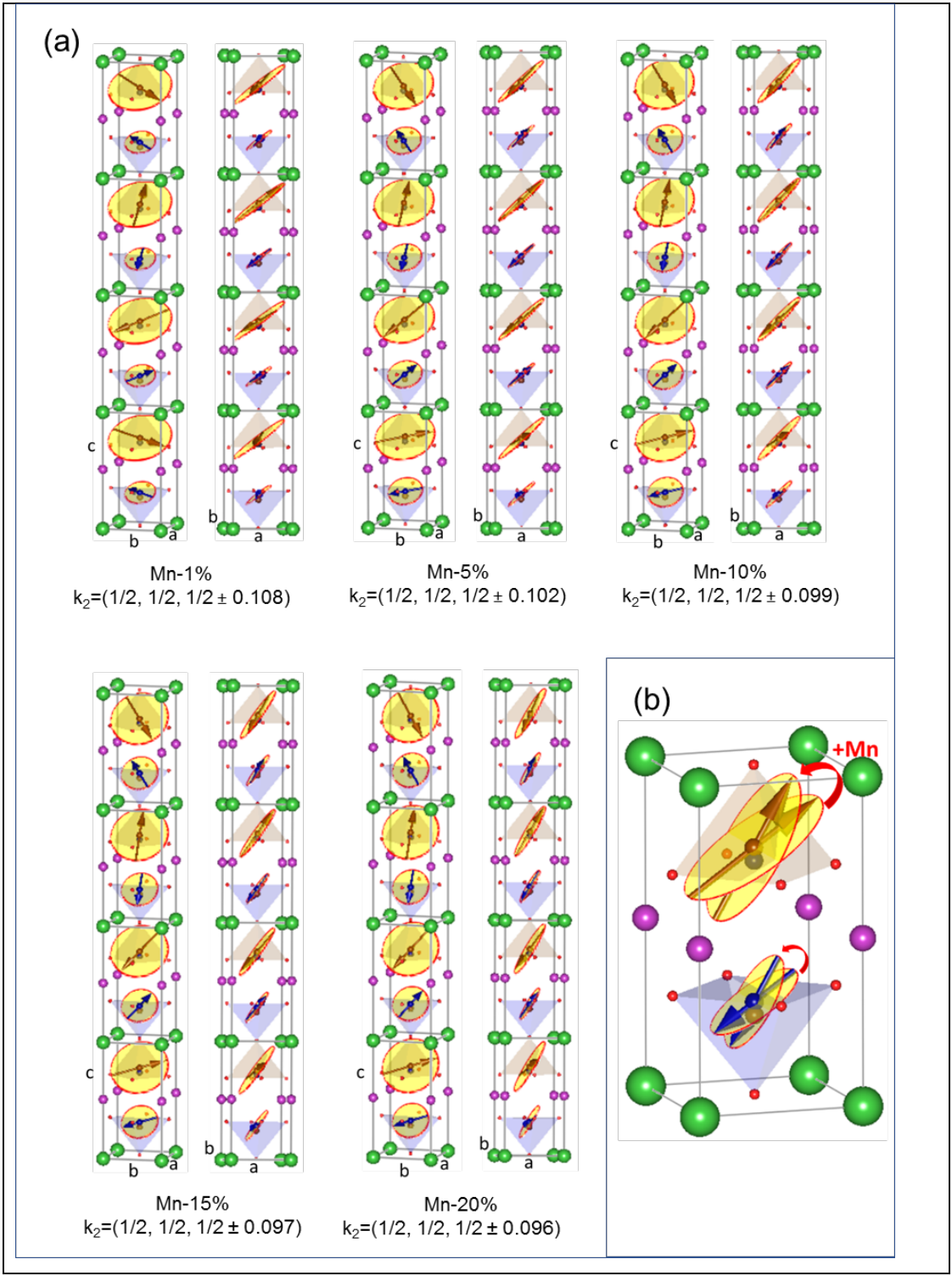

**Figure 10.** (a) Two projections of the refined ICM spiral order at 10K (with Mr=Mi, as described in Table II) in YBaCuFe$_{1-x}$Mn$_x$O$_5$ samples with x=1, 5, 10, 15 and 20% Mn. (b) Projection of the evolution of the tilting of the spiral rotation plane ($\theta$) in the ICM phase by increasing the Mn content. For clarity, the average magnetic moment is depicted in each pyramid (shared by three distinct metals) with the color of the majority metal.

In Figure 11 we have represented the dependence of the tilting angle $\theta$ in the CM collinear phase (at 300 K) and in the ICM regime (at 10K) with the amount of Mn. Hereafter we will associate the direction fixed by the Theta angle ($\theta$) as the easy-axis of the system for each magnetic phase (although the tetragonal symmetry does not allow identifying the direction within the **ab-**plane). We find that the orientation (tilting) of the magnetic easy-axis in the ICM phase differs from the CM collinear phase because $\theta_{ICM} < \theta_{CM}$. Moreover, this difference systematically increases with x, being a measure of the previously mentioned drop at the CM-ICM transition. Most interestingly, the evolution shown in Fig. 11 indicates that Mn substitution brings about a systematic reorientation of the plane of the helix. In such a way, the rotation plane tends to move away from the tetragonal **ab** plane approaching the **c** axis. Such evolution appears also illustrated in Figure 10(b). The observed rate is $\Delta\theta/\Delta x \approx -1.33$ deg/%Mn (Fig. 11). So, comparing the samples with x=0 and x=0.15 one can see a reorientation of the spiral plane (easy axis) of $\Delta\theta \sim -20°$, from 55° to 35°. We must stress upon the fact that here the orientation of the plane of the helix respect to the **c** axis determines if the spiral order is of cycloid- ($\theta \rightarrow 0°$) or helix-type ($\theta \rightarrow 90°$). Consequently, the ICM magnetic phase in YBaCuFe$_{1-x}$Mn$_x$O$_5$ would evolve from an inclined helix towards a cycloid by increasing the Mn content (see Figure 10(b)). It is thus important to recall at this point that for an electrical polarization **P** with origin at the antisymmetric Dzyaloshinskii–Moriya interaction, its dependence with the orientation of the helix in this system would be described by **P** $\alpha$ **r**$_{12}$ x (**S**$_1$ x **S**$_2$) (eq. [2]), being in this case **r**$_{12}$ the **c** axis and **S**$_1$ and **S**$_2$ the spins at Cu and Fe sites of a bilayer [12,13]. In the non-collinear case the spin-chirality **Q**=**S**$_1$x**S**$_2 \neq 0$, and the module of the polarization would be P $\alpha$ Q · $sin(90-\theta)$= Q · $\cos(\theta)$, namely, proportional to $\cos(\theta)$.

We performed additional refinements to study how the obtained $\theta$ angle changes with the constraint that relates Fe and Cu magnetic moments. As shown in Fig. 11, by using m(Fe)=m(Cu) $\theta$ turns out to be systematically shifted down by $\approx 12°$. The Rietveld refinements of the neutron diffraction patterns at 10K for all Mn doped samples are exposed in Fig. S3 of the Supplementary Information.

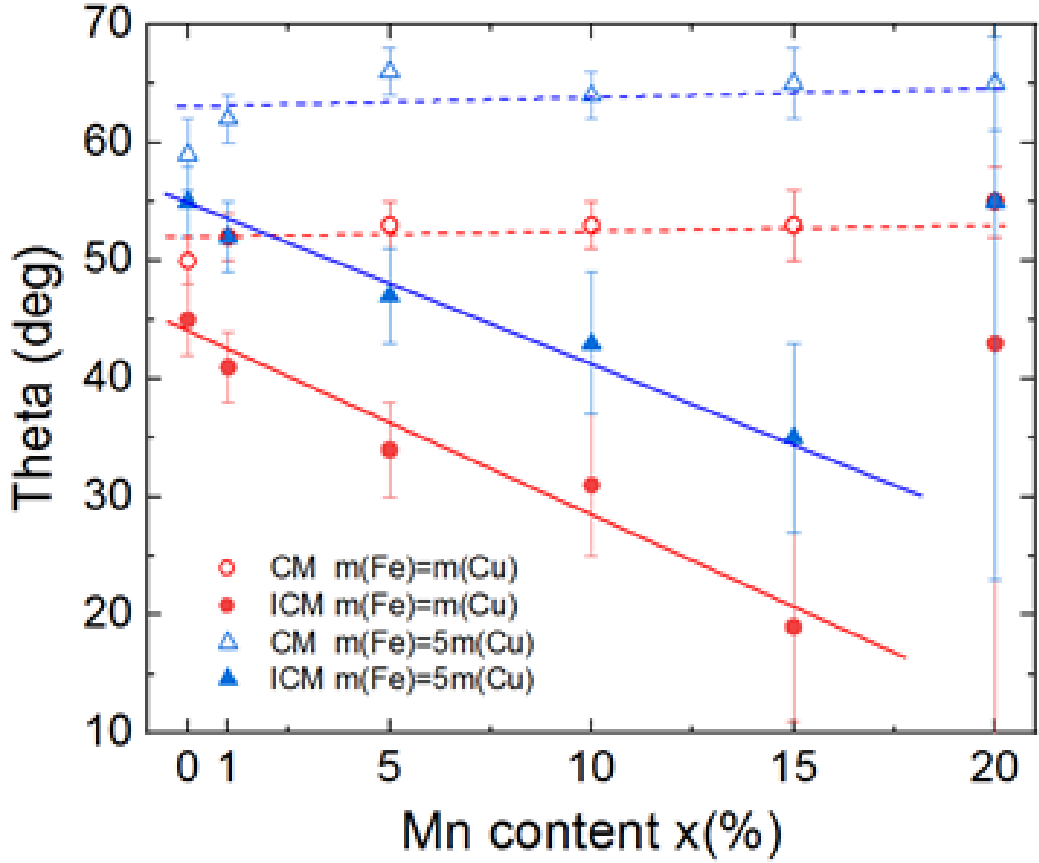

**Figure 11.** Inclination angle (θ) in the CM collinear (300 K, empty symbols) and ICM spiral phases (10 K, full symbols) in YBaCuFe$_{1-x}$Mn$_x$O$_5$ as a function of Mn content. The plot compares the refined values using two models (constraints): (Blue) m(Fe)=5m(Cu) and (Red) m(Fe)=m(Cu). (A nearly constant shift is observed between both).

## 4. Conclusions.

Contradictory results on the ferroelectric response of YBaCuFeO$_5$ type compounds are opening up a lively debate. On one hand, high T$_S$ transition temperatures are observed in RBaCuFeO$_5$ and YBa$_{1-x}$Sr$_x$CuFeO$_5$ samples studied in polycrystalline form. On the other hand, spontaneous polarization has not been reported in samples with Ts ≥300 K nor in single crystals. We have shown that a proper determination of the Mr/Mi ratio (important for ferroelectricity) requires quality single crystal neutron data. The hitherto absence of ferroelectricity in YBCFO crystals could be attributed to issues related with the chemical disorder or/and to specific features of the spiral (Mr/Mi ratio or inclination of the spiral rotation plane, for instance). In a recent experiment by Lai *et al.* [9] on YBCFO single crystal samples they observed null polarization and neutron powder diffraction intensities compatible with a helical spiral (parallel to the *ab* plane). Additionally, the strength of the spin-lattice interaction and the relevance of the DM mechanism is still uncertain in these oxides.

Within this context, in this work we have tried to increase the spin-orbit coupling effects acting on the B sites of the perovskite. We have synthesized and studied the family of $YBaCuFe_{1-x}Mn_xO_5$ perovskites where which trivalent Fe has been partially substituted by Mn. The evolution of their magnetic properties while increasing the Mn content have been investigated and compared to the undoped candidate to high-temperature spiral multiferroic YBCFO. This class of AA'BB'O$_5$ perovskites composed of bipyramidal layers exhibit complete order of A and A' cations but only partial order of B and B'. Since B-site ordering is sensitive to the cooling process during the last annealing, all compositions were prepared under identical conditions and a cooling rate of 300K/h.

The evolution of the upper and lower pyramids when increasing the presence of Mn atoms indicates that the Jahn-Teller splitting around Cu is preserved along the series. On another hand, the observed evolution rules out here the Jahn-Teller configuration that $Mn^{3+}$ ions adopt in the charge-order phase of $YBaMn_2O_5$ [20]. Although additional spectroscopic studies are needed, very likely this occurs due to a too-high energy penalty for the occurrence of both Cu and Mn atoms with occupied $3d$ $d_z^2$ orbitals parallel to the **c** axis in $CuMnO_9$ units.

The T-x magnetic phase diagram for $YBaCuFe_{1-x}Mn_xO_5$ has been built up to the paramagnetic state from NPD data and the evolution of the commensurate and incommensurate magnetic orders have been thoroughly described as a function of temperature and doping level. The evolution of Ts can be understood in the light of the theoretical model developed by Scaramucci *et al*. [7,8], mainly due to changes in the exchange interactions. The study of the spiral incommensurate order has been expanded to the opposite scenarios of very high (Mr>>Mi) and zero eccentricity (Mr=Mi).

Finally, two additional observations have to be highlighted. First, we have shown that (whilst keeping an invariant cooling rate) the degree of chemical disorder between Cu/Fe sites can be significantly raised by addition of a third (B'') cation. We have found that the fraction of B cations (Cu) that occupy B' positions (Fe site) notably increases with the presence of a third metal (B''=Mn) sharing B' positions in the *P4mm* structure. This finding can be applied as an alternative method to enhance the degree of disorder in samples prepared in particular forms, such as single-crystals. Second, we have shown that in $YBaCuFe_{1-x}Mn_xO_5$ neither the orientation of magnetic moments in the collinear phase nor the spin rotation plane in the spiral phase evolve or change with temperature. In other words, the tilting angles $\theta_{col}$ (=$\theta_{CM}$) and $\theta_{spiral}$ (=$\theta_{ICM}$) adopt constant values for a given composition. Within the error $\theta_{col}$ shows no dependence on the Mn content. Conversely and more interesting, $\theta_{spiral}$ exhibits a strong dependence on the Mn substitution. So, neutron diffraction have revealed a systematic reorientation of the plane of the helix ($\theta_{spiral}$): from rather perpendicular to rather parallel to the *c* axis when increasing the Mn content. Therefore, in this structure the presence of Mn can transform a helical spin order (**k**//**Q**) into one of cycloidal type (**k**⊥**Q**), producing respectively null and finite electrical polarization in DM improper ferroelectrics. This is of particular importance for single crystals in view of their tendency to get the moments of the helix oriented parallel to the *ab* plane [9]. These observations, based on changes in the B sites are of interest for engineering and developing this family of potential high-temperature multiferroics.

**Acknowledgements.**

We acknowledge financial support from the Spanish Ministerio de Ciencia, Innovación y Universidades (MINCIU), through Project No. RTI2018-098537-B-C21, cofunded by ERDF from EU, and "Severo Ochoa" Programme for Centres of Excellence in R&D (FUNFUTURE (CEX2019-000917-S)). X.Z. was financially supported by China Scholarship Council (CSC) with No. 201706080017. X.Z's work was done as a part of the Ph.D program in Materials Science at Universitat Autònoma de Barcelona. We also acknowledge ALBA, ILL and D1B-CRG (MINECO) for provision of beam time (dois: 10.5291/ILL-DATA.CRG-2655, 10.5291/ILL-DATA.CRG-2478, 10.5291/ILL-DATA.CRG-2562).

# References

[1] S. Dong, J.M. Liu, S.W. Cheong, Z. Ren, Multiferroic materials and magnetoelectric physics: Symmetry, entanglement, excitation, and topology, Adv. Phys. 64 (2015) 519–626. https://doi.org/10.1080/00018732.2015.1114338.

[2] V. Caignaert, I. Mirebeau, F. Bourée, N. Nguyen, A. Ducouret, J.M. Greneche, B. Raveau, Crystal and magnetic structure of ybacufeo5, J. Solid State Chem. 114 (1995) 24–35. https://doi.org/10.1006/jssc.1995.1004.

[3] B. Kundys, A. Maignan, C. Simon, Multiferroicity with high-TC in ceramics of the YBaCuFeO 5 ordered perovskite, Appl. Phys. Lett. 94 (2009) 32–35. https://doi.org/10.1063/1.3086309.

[4] M. Morin, A. Scaramucci, M. Bartkowiak, E. Pomjakushina, G. Deng, D. Sheptyakov, L. Keller, J. Rodriguez-Carvajal, N.A. Spaldin, M. Kenzelmann, K. Conder, M. Medarde, Incommensurate magnetic structure, Fe/Cu chemical disorder and magnetic interactions in the high-temperature multiferroic YBaCuFeO5, Phys. Rev. B. 91 (2015) 064408. https://doi.org/10.1103/PhysRevB.91.064408.

[5] M. Morin, E. Canévet, A. Raynaud, M. Bartkowiak, D. Sheptyakov, V. Ban, M. Kenzelmann, E. Pomjakushina, K. Conder, M. Medarde, Tuning magnetic spirals beyond room temperature with chemical disorder, Nat. Commun. 7 (2016) 13758. https://doi.org/10.1038/ncomms13758.

[6] T. Shang, E. Canévet, M. Morin, D. Sheptyakov, M.T. Fernández-Díaz, E. Pomjakushina, M. Medarde, Design of magnetic spirals in layered perovskites: Extending the stability range far beyond room temperature, Sci. Adv. 4 (2018). https://doi.org/10.1126/sciadv.aau6386.

[7] A. Scaramucci, H. Shinaoka, M. V. Mostovoy, M. Müller, C. Mudry, M. Troyer, N.A. Spaldin, Multiferroic Magnetic Spirals Induced by Random Magnetic Exchanges, Phys. Rev. X. 8 (2018) 11005. https://doi.org/10.1103/PhysRevX.8.011005.

[8] A. Scaramucci, H. Shinaoka, M. V. Mostovoy, R. Lin, C. Mudry, M. Müller, Spiral order from orientationally correlated random bonds in classical X Y models , Phys. Rev. Res. 2 (2020) 1–23. https://doi.org/10.1103/physrevresearch.2.013273.

[9] Y.C. Lai, C.H. Du, C.H. Lai, Y.H. Liang, C.W. Wang, K.C. Rule, H.C. Wu, H.D. Yang, W.T. Chen, G.J. Shu, F.C. Chou, Magnetic ordering and dielectric relaxation in the double perovskite YBaCuFeO 5, J. Phys. Condens. Matter. 29 (2017) 145801 (8pp). https://doi.org/10.1088/1361-648X/aa5708.

[10] M. Pissas, Magnetic texturing due to the partial ordering of Fe+3 and Cu+2 in NdBaCuFeO5, J. Magn. Magn. Mater. 432 (2017) 224–230. https://doi.org/10.1016/j.jmmm.2017.01.083.

[11] H.W. Chen, Y.W. Chen, J.L. Kuo, Y.C. Lai, F.C. Chou, C.H. Du, H.L. Liu, Spin-charge-lattice coupling in YBaCuFeO 5 : Optical properties and first-principles calculations, Sci. Rep. 9 (2019) 3223. https://doi.org/10.1038/s41598-019-39031-

6.

[12]  H. Katsura, N. Nagaosa, A. V. Balatsky, Spin current and magnetoelectric effect in noncollinear magnets, Phys. Rev. Lett. 95 (2005) 1–4. https://doi.org/10.1103/PhysRevLett.95.057205.

[13]  M. Mostovoy, Ferroelectricity in spiral magnets, Phys. Rev. Lett. 96 (2006) 1–4. https://doi.org/10.1103/PhysRevLett.96.067601.

[14]  D. Dey, S. Nandy, T. Maitra, C.S. Yadav, A. Taraphder, Nature of spiral state and absence of electric polarisation in Sr-doped YBaCuFeO 5 revealed by first-principle study, Sci. Rep. 8 (2018) 1–9. https://doi.org/10.1038/s41598-018-20774-7.

[15]  S. Lal, S.K. Upadhyay, K. Mukherjee, C.S. Yadav, Evolution of magnetic and dielectric properties in Sr-substituted high-temperature multiferroic YBaCuFeO5, Epl. 117 (2017) 67006. https://doi.org/10.1209/0295-5075/117/67006.

[16]  R.D. Shannon, Revised Effective Ionic Radii and Systematic Study of Inter Atomic Distances in Halides and Chalcogenides in Halides and Chaleogenides, (2016). https://doi.org/10.1107/s0567739476001551.

[17]  F. Fauth, R. Boer, F. Gil-Ortiz, C. Popescu, O. Vallcorba, I. Peral, D. Fullà, J. Benach, J. Juanhuix, The crystallography stations at the Alba synchrotron, Eur. Phys. J. Plus. (2015). https://doi.org/10.1140/epjp/i2015-15160-y.

[18]  J. Rodríguez-Carvajal, Recent advances in magnetic structure determination by neutron powder diffraction, Phys. B Phys. Condens. Matter. 192 (1993) 55–69. https://doi.org/10.1016/0921-4526(93)90108-I.

[19]  W.C. Liu, Y.Z. Zheng, Y.C. Chih, Y.C. Lai, Y.W. Tsai, Y.Z. Zheng, C.H. Du, F.C. Chou, Y.L. Soo, S.L. Chang, X-ray multi-beam resonant diffraction analysis of crystal symmetry for layered perovskite YBaCuFeO5, J. Appl. Crystallogr. 49 (2016) 1721–1725. https://doi.org/10.1107/S1600576716013248.

[20]  F. Millange, E. Suard, V. Caignaert, B. Raveau, YBaMn2O5: Crystal and magnetic structure reinvestigation, Mater. Res. Bull. 34 (1999) 1–9. https://doi.org/10.1016/S0025-5408(98)00214-1.

[21]  J.B. Goodenough, An interpretation of the magnetic properties of the perovskite-type mixed crystals La1-xSrxCoO3-λ, J. Phys. Chem. Solids. 6 (1958) 287–297. https://doi.org/10.1016/0022-3697(58)90107-0.

[22]  J. Kanamori, Superexchange interaction and symmetry properties of electron orbitals, J. Phys. Chem. Solids. 10 (1959) 87–98. https://doi.org/10.1016/0022-3697(59)90061-7.

[23]  P.W. Anderson, New approach to the theory of superexchange interactions, Phys. Rev. 115 (1959) 2–13. https://doi.org/10.1103/PhysRev.115.2.

# Supplementary information

## Tuning the tilting of the spiral plane by Mn doping in YBaCuFeO$_5$ multiferroic

X. Zhang[1], A. Romaguera[1], O. Fabelo[2], F. Fauth[3], J. Herrero-Martín[3] and J.L. García-Muñoz[1*]

[1]Institut de Ciència de Materials de Barcelona, ICMAB-CSIC, Campus UAB, 08193 Bellaterra, Spain
[2]ILL-Institut Laue Langevin, 38042 Grenoble Cedex, France.
[3]ALBA-CELLS Synchrotron Light Source, 08290 Cerdanyola del Vallès, Barcelona, Spain.

This PDF file includes:

**Fig. S1**. Refinement of the D2B neutron pattern for the YBCFO sample.

**Fig. S2**. Evolution of the cell parameters.

**Fig. S3**. Rietveld fits of neutron patterns at 10 K for Mn doped samples.

**Table S1**. Structural results of the Rietveld fits and interatomic distances for the YBCFO sample of this work (from high-resolution SXRD and NPD data) and for the YBCFO sample in PRB 91, 064408 (2015) (NPD).

**Table S2**. Interatomic distances in YBaCuFe$_{1-x}$Mn$_x$O$_5$.

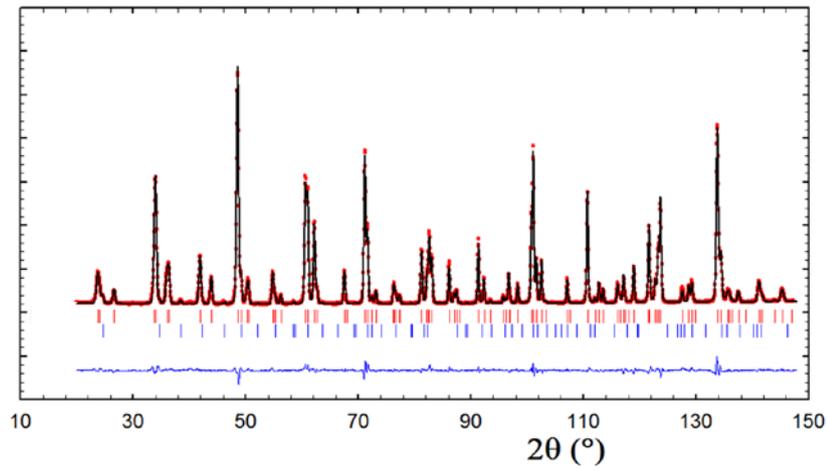

**Figure S1.** Rietveld refinement (black curve) of the D2B neutron pattern (red circles) at 300 K for the YBCFO sample. The second row of bars correspond to the collinear magnetic phase.

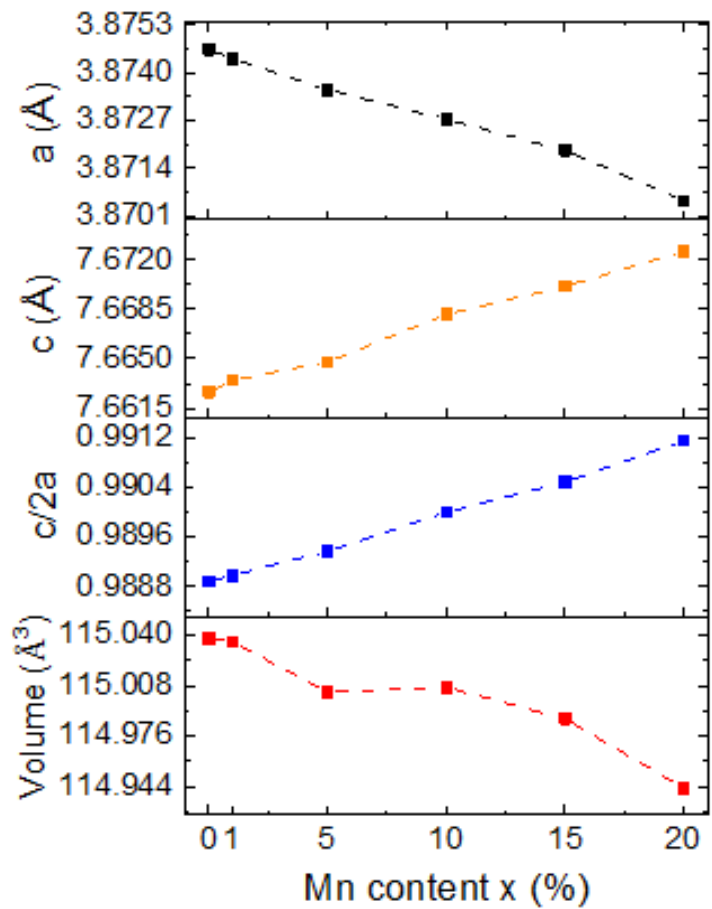

**Figure S2.** YBaCuFe$_{1-x}$Mn$_x$O$_5$ : evolution with Mn doping of the cell parameters and volume from SXPD data at room temperature.

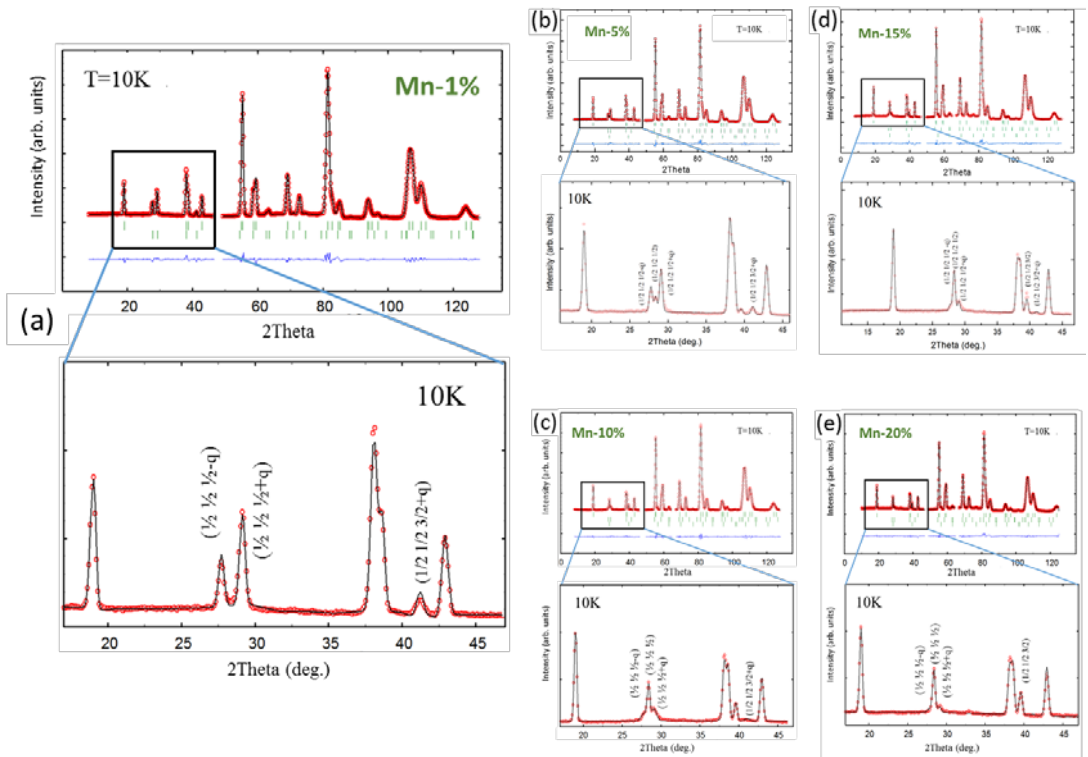

**Figure S3.** Rietveld analysis of the neutron diffraction patterns at 10 K (d1b@ILL) for **(a) Mn-1%** ($R_B$:1.67, $R_f$:0.94, $R_M$:6.9, $\chi^2$:12.3) ; **(b) Mn-5%** ($R_B$:2.12, $R_f$:1.22, $R_M$:15.6, $\chi^2$:13.2); **(c) Mn-10%** ($R_B$: 0.98, $R_f$:0.57, $R_M$:12.9, $\chi^2$:7.6); **(d) Mn-15%** ($R_B$: 1.86, $R_f$:1.04, $R_M$:23.3, $\chi^2$:10.2); **(e) Mn-20%** ($R_B$: 2.25, $R_f$:1.29, $R_M$:43.0, $\chi^2$:12.3). Rows of bars correspond to allowed reflections for the structural, ICM magnetic (circular spiral) and CM magnetic phases. The expanded low-angle region is also shown for each composition, refined using the average moment [m(Fe)=m(Cu)].

**Table S1**. Refined structural parameters at 300 K (*P4mm*) and main interatomic distances (Å) for the same undoped (x=0) sample from SXRD and high-resolution NPD data. Synchrotron (X) and neutron (n) patterns were recorded respectively on mspd@Alba (λ=0.41338 Å, fig. 1) and d2b@ILL (λ= 1.594 Å, fig. S1). The structure for the YBCFO sample reported in Ref. [4] is also shown.

| $YBaCuFeO_5$ | MSPD (X) this work | D2B (n) this work | HRPT (n) Ref. [4]* |
|---|---|---|---|
| $a$ | 3.87463(3) | 3.87424(1) | 3.87325(1) |
| $c$ | 7.66267(5) | 7.66036(3) | 7.6655(3) |
| z (Y) (0 0 z) | 0.5097(5) | 0.5036(7) | 0.5053(16) |
| z (Ba) (0 0 z) | 0 | -0.0112(4) | 0 |
| z (Cu1) (½ ½ z) | 0.7145(3) | 0.7149(4) | 0.7144(7) |
| z (Fe1) (½ ½ z) | 0.7540(6) | 0.7466(4) | 0.7484(8) |
| z (Cu2) (½ ½ z) | 0.2855(3) | 0.2851(4) | 0.2856(7) |
| z (Fe2) (½ ½ z) | 0.2460(6) | 0.2534(4) | 0.2516(8) |
| z ($O_1$) (½ ½ z) | 0.010(3) | 0.0112(4) | 0.0179(15) |
| z ($O_2$) (0 ½ z) | 0.327(1) | 0.3233(6) | 0.3265(15) |
| z ($O_3$) (0 ½ z) | 0.693(1) | 0.6921(6) | 0.6947(15) |
| Occ (Cu2 and Fe1) | 0.772(22) | 0.624(20) | 0.703(2) |
| $\chi^2$ | 50.0 | 8.93 | 1.98 |
| $R_B$ | 6.02 | 2.50 | 3.81 |
| $R_f$ | 7.28 | 1.64 | 3.86 |
| d(Fe-$O_a$) | 1.962(16) | 2.027(4) | 2.066 |
| d(Fe-$O_b$) | 1.992(4) | 1.982(1) | 1.980 |
| d(Cu-$O_a$) | 2.107(16) | 2.098(4) | 2.052 |
| d(Cu-$O_b$) | 1.963(4) | 1.959(1) | 1.962 |
| d(Fe-Cu) | 0.303 (5) | 0.247 (4) | 0.261 |

*Ref. [4]: Morin *et al.* Phys. Rev. B. 91, 064408 (2015). The distances are calculated from the reported structure.

**Table S2.** Refined interatomic distances (in Angstroms) at 300 K (*P4mm*) in layered YBaCuFe$_{1-x}$Mn$_x$O$_5$ from SXRD data (mspd@Alba).

| distances (Å) | x = 0 | x = 0.01 | x = 0.05 | x = 0.10 | x = 0.15 | x = 0.20 |
|---|---|---|---|---|---|---|
| *a = b* | 3.87463 (2) | 3.87439 (2) | 3.87354 (3) | 3.87274 (4) | 3.87189 (4) | 3.87052 (4) |
| *c* | 7.66267 (5) | 7.66349 (5) | 7.66479 (7) | 7.66816 (8) | 7.67013 (8) | 7.67265 (8) |
| d(Fe-O$_a$) | 1.962 (16) | 1.973 (20) | 1.964 (20) | 1.962 (20) | 1.953 (20) | 1.941 (20) |
| d(Fe-O$_b$) | 1.992 (4) | 1.988 (4) | 1.987 (4) | 1.986 (4) | 1.982 (4) | 1.984 (4) |
| d(Cu-O$_a$) | 2.107 (16) | 2.117 (20) | 2.098 (20) | 2.107 (20) | 2.091 (20) | 2.119 (20) |
| d(Cu-O$_b$) | 1.963 (4) | 1.962 (4) | 1.963 (4) | 1.960 (4) | 1.964 (4) | 1.955 (4) |
| d(Fe-Cu) | 0.303 (5) | 0.269 (5) | 0.307 (5) | 0.295 (5) | 0.275 (5) | 0.278 (5) |